\documentclass[preprint2,longabstract]{aastex}

\slugcomment{Based on observations collected at the European Southern Observatory, Chile, through proposal 381.C-0860(A).}

\shorttitle{NIR spectroscopy and modeling of Eps Indi B}
\shortauthors{Kasper et al.}

\usepackage{natbib}

\begin{document}


\title{Testing the models: NIR imaging and spectroscopy of the benchmark T-dwarf binary Eps Indi B}


\author{M. Kasper}
\affil{European Southern Observatory (ESO), Karl-Schwarzschild-Str. 2, 85748 Garching, Germany}
\email{mkasper@eso.org}

\author{A. Burrows}
\affil{Princeton University, Department of Astrophysical Sciences, Princeton, NJ 08544-1001, USA}

\and

\author{W. Brandner}
\affil{Max-Planck-Institut f\"ur Astronomie, K\"onigstuhl 17, 69117 Heidelberg, Germany}




\begin{abstract}
The relative roles of metallicity and surface gravity on the near-infrared spectra of late-T brown dwarfs are not yet fully understood, and evolutionary models still need to be calibrated in order to provide accurate estimates of brown dwarf physical parameters from measured spectra. The T-type brown dwarfs Eps Indi Ba and Bb forming the tightly bound binary Eps Indi B, which orbits the K4V star Eps Indi A, are nowadays the only such benchmark T dwarfs for which all important physical parameters such as metallicity, age and mass are (or soon will be) known. 
We present spatially resolved VLT/NACO images and low resolution spectra of Eps Indi B in the J, H and K near-infrared bands. The spectral types of Eps Indi Ba and Bb are determined by direct comparison of the flux-calibrated JHK spectra with T dwarf standard template spectra and also by NIR spectral indices. Eps Indi Bb is confirmed as a T6 while the spectral type of Eps Indi Ba is T1.5 so somewhat later than the previously reported T1. Constrained values for surface gravity and effective temperature are derived by comparison with model spectra. The evolutionary models predict masses around $\sim$53 M$_{\rm J}$ for Eps Indi Ba and $\sim$34 M$_{\rm J}$ for Eps Indi Bb, slightly higher than previously reported values. The suppressed J-band and enhanced K-band flux of Eps Indi Ba indicates that a noticeable cloud layer is still present in a T1.5 dwarf while no clouds are needed to model the spectrum of Eps Indi Bb.
\end{abstract}





\section{Introduction}

Multiple stellar systems with one component being a main sequence star and at least one other component having a mass below the hydrogen-burning limit, and hence residing in the domain of brown dwarfs are the key to calibrate and benchmark brown dwarf evolution models. Since all components are expected to be coeval and of the same chemical composition, age and metallicity of the main sequence primary determined by well established methods also apply to the brown dwarf companion(s). Furthermore, sufficiently tight binary systems would allow one to directly measure the system mass by determining the orbit and even to measure the individual component masses via monitoring of radial velocities. Hence, the measured spectral energy distribution (SED) of such a companion can be used to calibrate the effective temperature and surface gravity of evolutionary models and provide the required benchmarks. 

One such precious object, Eps Indi B, was resolved into a binary T1/T6 dwarf \citep{mccaughrean04} orbiting the K4.5V primary Eps Indi A during the commissioning of the Simultaneous Differential Imager \citep[SDI,][]{close05} at the VLT. Initial mass estimates based on spectral properties and an assumed age of 1.3\,Gyr were $47\pm10$M$_{\rm Jup}$ and $28\pm7$M$_{\rm Jup}$ for Eps Ind Ba and Bb, respectively \citep{mccaughrean04}.
Another T6 dwarf companion, SCR 1845 B, was recently found by \citet{biller06} around the M8.5 primary SCR 1845 A using the same instrument and characterized by NIR imaging and spectroscopy by \citet{kasper07}. Both, Eps Indi B and SCR 1845 B, are extremely nearby (closer than 4 pc), so near that they can efficiently be observed with modern Adaptive Optics instruments despite their intrinsic faintness and rather small orbital distance from the primary. 
Other benchmark objects include HN Peg \citep{leggett08}, Gl 570D \citep{saumon06} and HD 130948 BC \citep{dupuy08,potter02,goto02}.

It is now widely accepted that, besides T$_{\rm eff}$ (a function of bolometric luminosity and radius which are both a function of age), surface gravity (a function of mass and radius) and metallicity are centrally important in determining the spectral energy distribution of late-T dwarfs \citep{saumon07,liu07}. The effects of metallicity and surface gravity on the NIR spectra are, however, not yet fully understood \citep{burgasser06b, liu07}. Calibration of theoretical models against observations has started with the work by \citet{burgasser06b,saumon06,leggett07}. 
Both mass and age, i.e., the main underlying physical parameters for any modelling of brown dwarf evolution and atmospheres, are, however, still highly uncertain for most known T dwarfs. 

Precise mass estimates can only be derived for brown dwarf, which are members of binary or multiple systems, and for which orbital parameters have been determined. Currently, Eps Indi Bb is the only late T dwarf for which physical parameters, and in particular its mass,  will soon be known with high precision. While the Eps Indi B binary is separated from the Eps Indi A by 1500 AU (several arcminutes on sky at the distance of 3.626\,pc), the two brown dwarfs are separated only by about 0.7\arcsec{} corresponding to a projected spatial separation of 2.6 AU \citep{mccaughrean04}.  This suggests that the system mass might be measured from orbital motion within a few years. From measurements on Eps Indi A, the system age is reasonably well determined to be in the range 0.8 to 2 Gyr \citep{lachaume99}, and a near-to-solar metallicity ([Mg/H] = -0.05,[Al/H] = -0.04, [Fe/H] = -0.06) is reported by \citet{beirao05}.

\section{Observations and Data Reduction}

\begin{figure}
   \centering
   \includegraphics[width=\columnwidth]{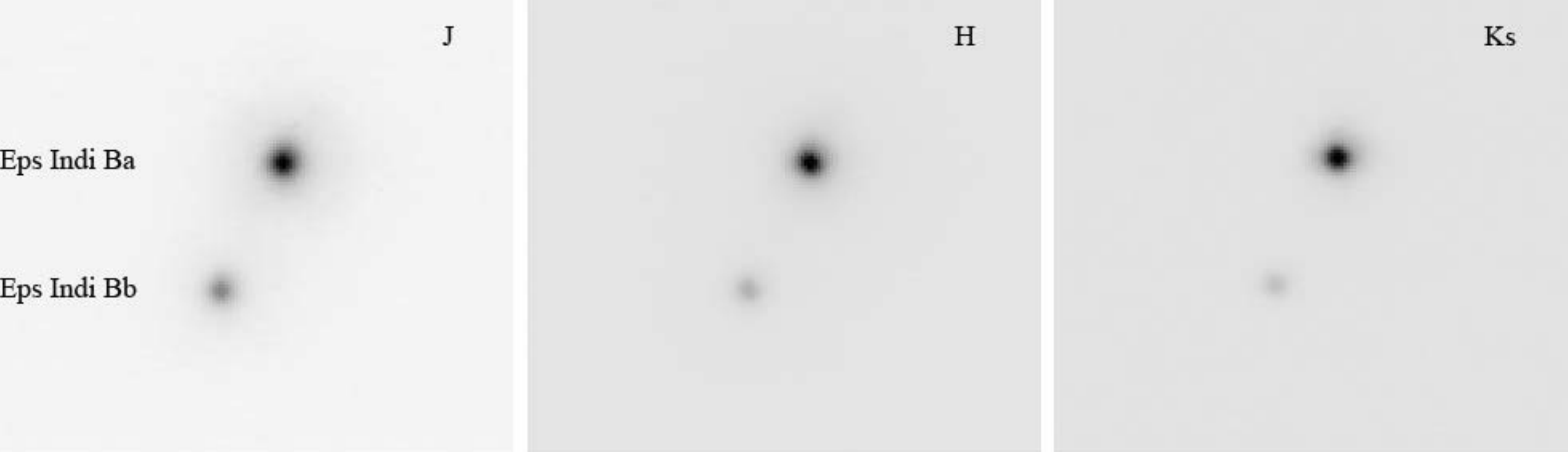}
   \caption{NACO broad band JHK images of the 0\farcs583 binary Eps Indi Bb. North is up, East is left.}
   \label{fig:jhkima}
\end{figure}

Eps Indi B was observed with NACO mounted on UT4 of the VLT on 23 May 2008 (imaging) and 9 June 2008 (spectroscopy). The Adaptive Optics used 90\% of the light from the binary as a guide star for its infrared wavefront sensor while 10\% of the light were directed towards the scientific camera. In this way, we obtained well corrected images that easily resolve the 0.58\arcsec{} binary (see Figure\,\ref{fig:jhkima}).

\subsection{Imaging}
For the imaging observations, the S13 camera of NACO was used in the J, H- and Ks near-infrared filters, with a pixel scale of $13.20\pm0.05$\,mas per pixel. The target was observed using a standard dithering procedure integrating for three minutes on source in each of the bands.

Image data reduction followed standard procedures including sky removal, bad pixel cleaning, and flat fielding. Differential photometry and astrometry were done on the reduced images by simultaneously fitting a model PSF profile to the binary. The individual magnitudes for Eps Indi Ba and Bb were finally calculated from the relative magnitudes and the integrated 2MASS magnitudes of the unresolved system (J=11.91, H=11.31 and Ks=11.21). The error introduced by the difference between NACO and 2MASS instrument responses has been calculated from the spectra, as described in the next section, and has been taken into account. Table\,\ref{tab:photastrom} displays 2MASS photometry of Eps Indi Ba and Bb. On 23 May 2008, the separation between the two components was 0\farcs583$\pm$0\farcs003, and the position angle was 153$\fdg$9$\pm$0$\fdg$5. The errors of the astrometric positions and the flux are the 95\% confidence levels provided by the MATLAB fitting algorithm, and the error on position angle is introduced by the uncertainty of the NACO FoV orientation.

\begin{table}
\begin{center}
\caption{2MASS photometry of Eps Indi Ba and Bb from 23 May 2008 (JD 2454609.9).}
\label{tab:photastrom}
\begin{tabular}{llll}
\tableline\tableline
Star & J & H & Ks \\
\tableline
Ba & 12.29$\pm$0.03 & 11.52$\pm$0.03 & 11.36$\pm$0.03 \\
Bb & 13.23$\pm$0.03 & 13.21$\pm$0.03 & 13.47$\pm$0.03 \\ 
\tableline
\end{tabular}
\end{center}
\end{table}

\subsection{Spectroscopy}
NACO was used in long-slit spectroscopic mode to obtain $R\approx400$ spectra in the J and HK spectroscopic filters (modes S54-4-SJ and S54-4-SHK, 2nm/px pixel dispersion, S54 camera with 54.3 mas/pixel). By turning the Nasmyth adaptor rotator, the slit was aligned with the binary axis. Dithering the binary at two positions along the slit, exposures of 26 minutes total in each of the modes were obtained. A telluric calibrator (HIP 113982, spectral type A0V) has been observed in a similar manner just after the object at a similar airmass. The effects of differential slit losses between Eps Indi B and the spectral calibrator \citep{goto03} were mitigated by observing both properly centered on the slit with about the same orientation of the parallactic angle with respect to the dispersion direction. A spectrum of the NACO internal Tungsten lamp was recorded for wavelength calibration.

Spectroscopic data reduction followed standard procedures. After sky removal, bad pixel cleaning and flat fielding, the spectra from Eps Indi Ba and Bb were extracted and wavelength calibrated. The halo of Eps Indi Ba at the position of Bb was estimated at its opposite position with respect to Eps Indi Ba. The spectra were divided by the telluric calibrator and multiplied by a Kurucz template spectrum of Vega (http://kurucz.harvard.edu/stars/VEGA/) smoothed to the actual resolution of the data in order to remove the hydrogen absorption features of the A0V telluric calibrator. Synthetic magnitudes were calculated as described in \citet{kasper07} in order to determine and apply the correction factors needed to convert the flux ratios from the NACO to the 2MASS photometric system. Flux calibration of the spectra was finally established using the calculated 2MASS magnitudes of both components (see table\,\ref{tab:photastrom}) and the 2MASS zero magnitude in-band flux \citep{cohen03}. 

It was verified that the errors introduced by the difference of the isophotal wavelengths for Eps Indi B and and the spectral calibrator do not exceed the photometric errors. We also estimated noise from the standard deviation of the individual spectra that were generated during the data analysis and averaged to obtain the final result. The SNR for each individual data point is around 30 for Eps Indi Ba and most parts of Eps Indi Bb spectra. In the K-band and in the depression between the J-band peaks, the SNR of the Eps Indi Bb spectrum drops to ~3-10. The deduced NIR spectrum is consistent with the optical to mid-IR spectra presented by \citet{king08}.

\section{Results}

\subsection{Spectral Types}
\begin{figure}
\centering
\includegraphics[width=\columnwidth]{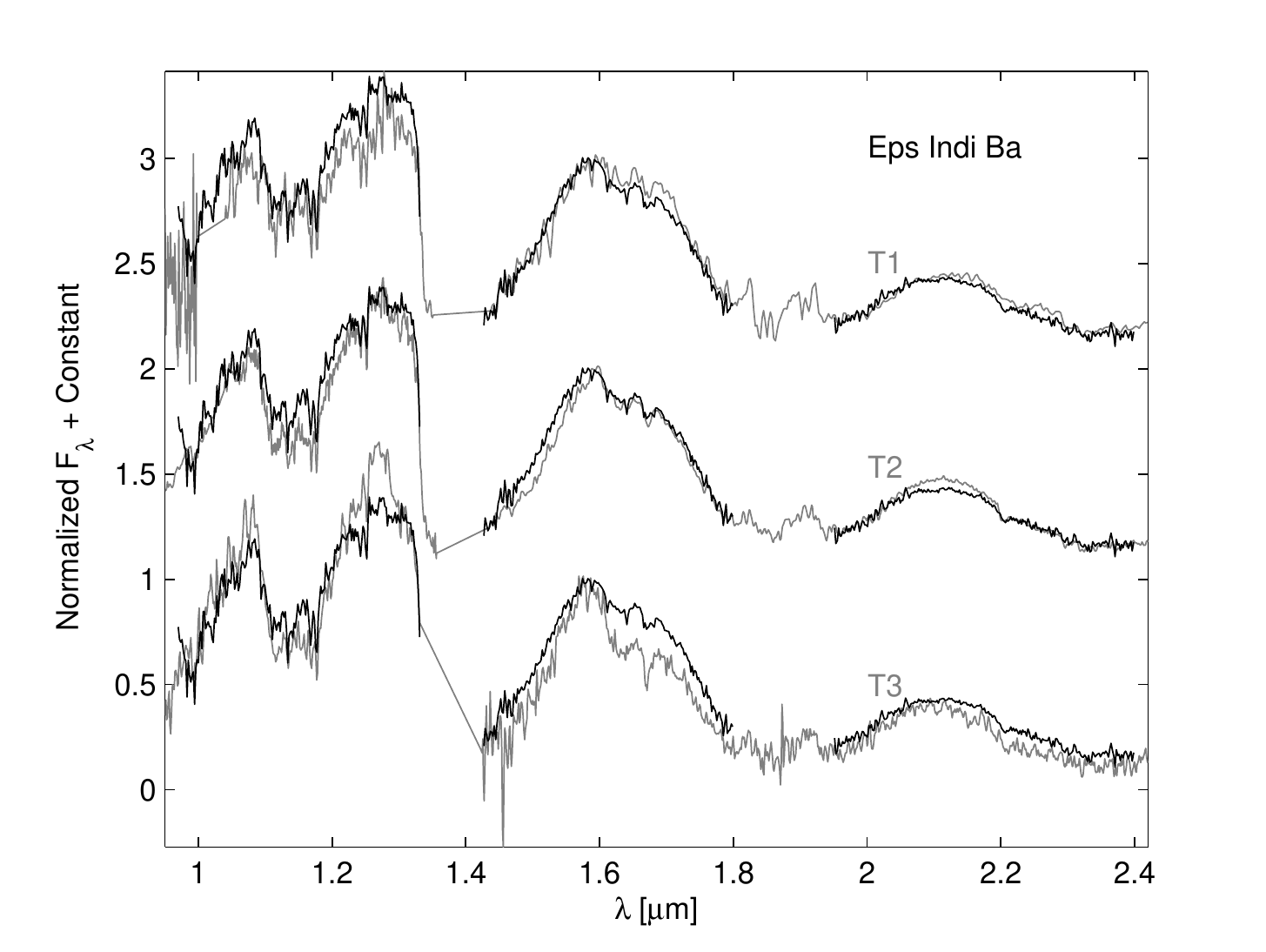}
\caption{Spectrum of Eps Indi Ba (black line) plotted on top of T1-3 dwarf spectral standards obtained from Sandy Leggett's L and T dwarf data archive.}
\label{fig:SpTBa}
\end{figure}

\begin{figure}
\centering
\includegraphics[width=\columnwidth]{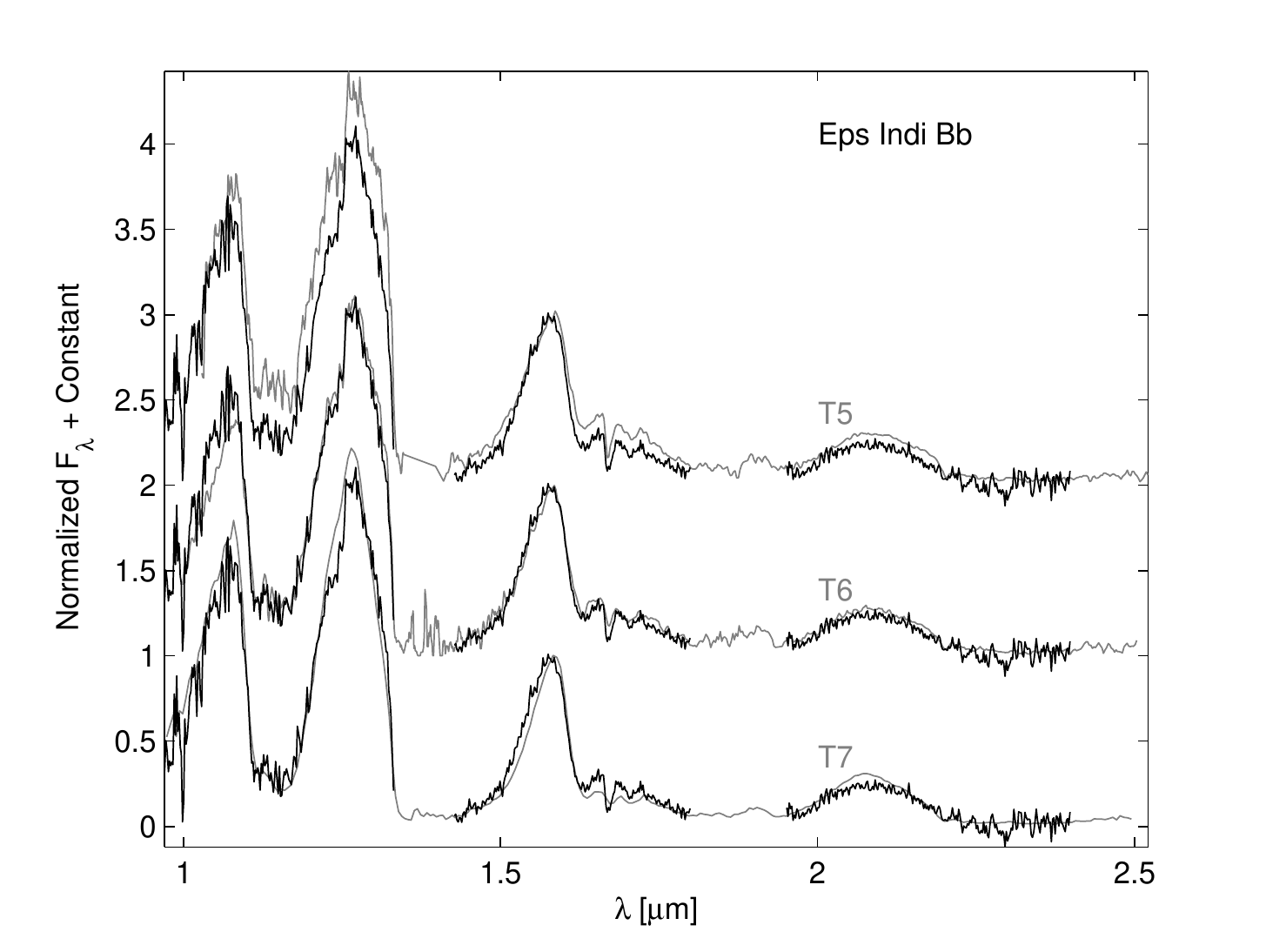}
\caption{Spectrum of Eps Indi Bb (black line) plotted on top of T5-7 dwarf spectral standards obtained from Sandy Leggett's L and T dwarf data archive.}
\label{fig:SpTBb}
\end{figure}

The most direct way to determine the spectral type of a star is direct comparison to spectral standards. Figures\,\ref{fig:SpTBa} and \ref{fig:SpTBb} plot the NIR spectrum of Eps Indi Ba and Bb (black lines) on top of T dwarf spectra obtained by 
\citet{knapp04,strauss99,leggett00, chiu06} and defined as spectral standards by \citet{burgasser06a}. The NIR spectrum of Eps Indi Ba fits somewhat in between the standard T1 and T2 spectra. Using the spectral indices defined by \citet{burgasser06a}, three (H$_2$O-J, H$_2$O-H, CH$_4$-K) suggest T1 while CH$_4$-H suggests T2 and CH$_4$-J even T3. Hence, we slightly refine the spectral type of Eps Indi Ba to T1.5. The spectrum of Eps Indi Bb fits the T6 standard well.

\subsection{Fitting Models to the Eps Indi Ba+b Data}

Assuming the Eps Indi B components have the same metallicity (near solar, \citet{beirao05}) and age ($\sim0.8 - 2$ Gyr, \citet{lachaume99}) allows us  to explore the accuracy of basic brown dwarf evolution and atmosphere theory.  Evolutionary theory provides the mapping between, on the one hand, mass, age, and radius and, on the other, T$_{\rm eff}$ and gravity ($g$) (for a given metallicity). For given T$_{\rm eff}$ and gravity, we have a predicted radius, and with the known distance we can then make {\it absolute} comparisons between the meaured spectra and theoretical models.  For the evolutionary theory, we use \citet{burrows97}. For the spectral and atmosphere theory, we use the approach described in \citet{burrows06} and \citet{sharp07}. The goal is to determine T$_{\rm eff}$ and gravity for each object, and then to determine whether a consistent theoretical picture emerges.  In particular, we can ascertain whether the models are consistent with the assumption of coevality.  

These current models have some limitations. First, since Eps Indi Ba is clearly near the L/T transition region above a T$_{\rm eff}$ of $\sim$1100 K, it is necessary to incorporate the effects of silicate clouds \citep{ackerman01, tsuji02, burrows06}.  The modeling of clouds (e.g., their modal particle sizes, spatial extent, optical properties, and degree of patchiness) is quite problematic.  Nevertheless, we use the "Case E" cloud model described in \citep{burrows06} with 15\,$\mu$m spherical  forsterite grains of which the cloud is assumed to be comprised. We initially also tested 10 and 20\,$\mu$m particles but did not achieve better fits. Second, though the absorption feature due to the hot bands of methane from $\sim$1.6 to $\sim$1.75\,$\mu$m is a defining signature of brown dwarfs, the strengths of these bands are not well-determined theoretically, and have not been measured in the laboratory.  Since it is as good as any credible formulation described in the extant literature and though it clearly underestimates the strength of these bands, we use the prescription described in \citet{sharp07}. Finally, the wings of the K I resonance absorption doublet at $\sim$0.77 $\mu$m determine the continuum of brown dwarf spectra out to as far as $\sim$1.05 $\mu$m \citep{burrows00}. The shape of these wings, in particular the effective cutoff wavelength, is still a subject of active investigation.

Given these caveats, we have identified the range of spectral models which approximately reproduce the observations. For Eps Indi Ba, we ran grids of models spanning the T$_{\rm eff}$ range 1150-1300\,K with a resolution of 50\,K and the log($g$\,[cm s$^{-2}$]) range 4.9-5.3 with a resolution of 0.1. For Eps Indi Bb, we ran grids of models spanning the T$_{\rm eff}$ range 850-950 K with a resolution of 25 K and the log($g$) range 4.7-5.1 with a resolution of 0.1. The models were than compared to the measured spectra by visual inspection looking for the smallest residuals between data and model over the complete spectral region. We further calculated least-squares residuals between data and model and found that the results agree very well with the by-eye fits. We did not attempt to use more sophisticated numerical methods such as $\chi^2$ statistics \citep{cushing08, burgasser07} to quantify the goodness-of-fit because current brown dwarf spectral theories have yet too many unquantified systematic problems that make a rigorous numerical treatment difficult and to some extent still arbitrary. Masses and ages were derived for the three best fitting models from the evolutionary tracks of \citet{burrows97} by linear interpolation. We then determined the final mass and age estimate by calculating a weighted average of the three best fitting models assigning weights of 1, 0.5 and 0.33 in order of preference. Errors on these final estimates were derived from the mass and age variations introduced by our $\log(g)$ step size of 0.1 which dominate over the variations introduced by the chosen T$_{\rm eff}$ resolution.

\begin{figure}
\centering
\includegraphics[width=\columnwidth]{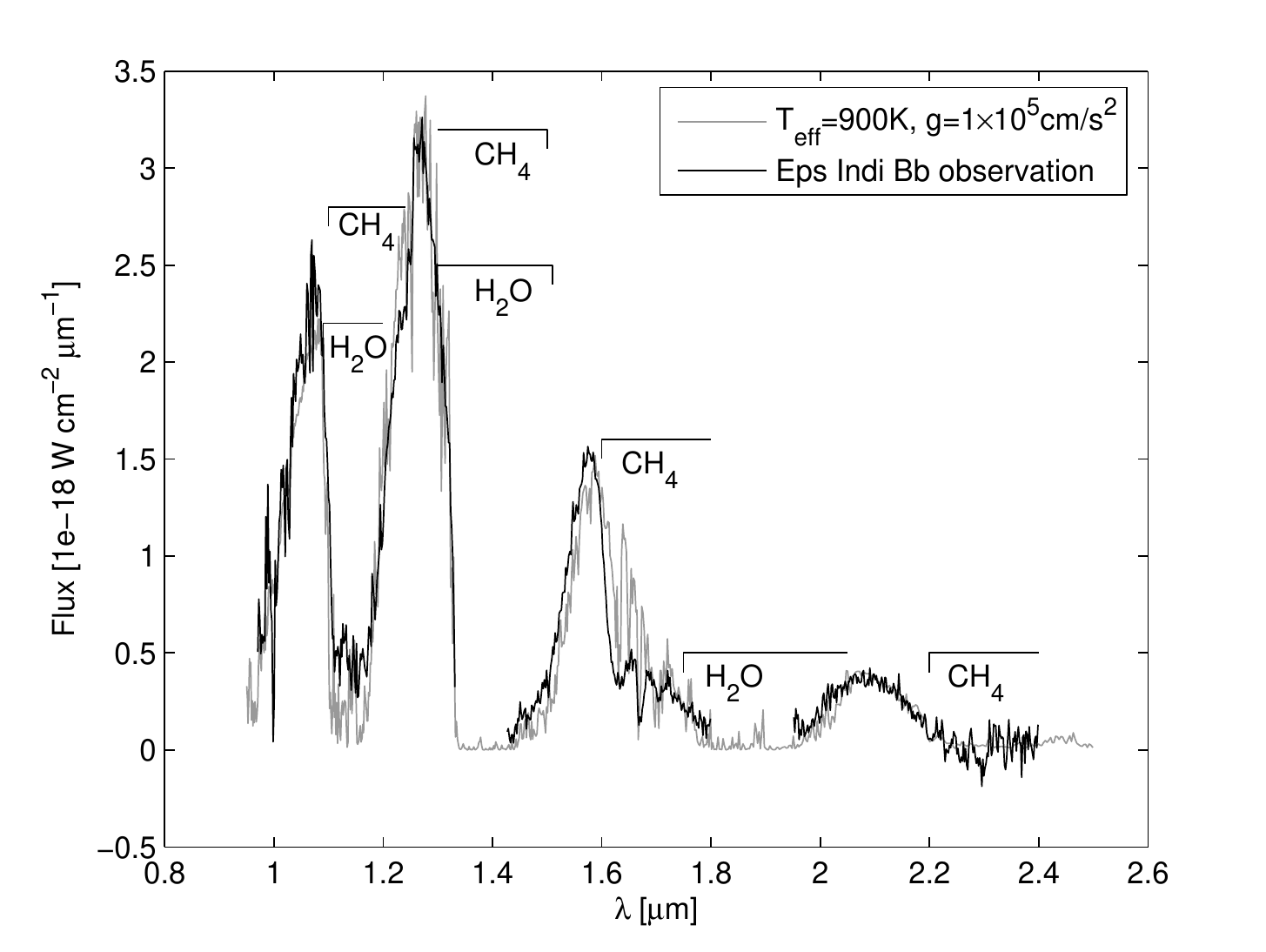}
\caption{NIR spectrum of Eps Indi Bb. The underlying gray curve shows the best model fit.}
\label{fig:jhkspectraB}
\end{figure}

For Eps Indi Bb, no cloud models were necessary and we derive well matching fits for (T$_{\rm eff}$ / log($g$)) of (900 / 5.0), (925 / 5.1) and (875 / 4.9) in order of preference. The best-fit values of T$_{\rm eff}$ and $g$ are correlated, with the higher gravities paired with the higher T$_{\rm eff}$s. Using the evolutionary tracks, the corresponding masses and ages are (34/,M$_{\rm J}$ / 1.5\,Gyr), (38\,M$_{\rm J}$ / 1.8\,Gyr) and (30\,M$_{\rm J}$ / 1.1\,Gyr). Using the weighted average described in the previous paragraph, we finally estimate mass and age of Eps Indi Bb of 34$^{\pm5}$\,M$_{\rm J}$ and 1.5$^{\pm0.5}$\,Gyr with the error bars derived from the variations introduced by the log($g$) resolution. The radius of Eps Indi Bb, taken from \citet{burrows97}, is $\sim$0.93 R$_{\rm J}$. Generally, the theoretical radii do not vary by more than $\sim$10\%, with lower radii associated with higher values of $g$. Figure\,\ref{fig:jhkspectraB} displays the measured spectrum together with the best fitting model. 

\begin{figure}
\centering
\includegraphics[width=\columnwidth]{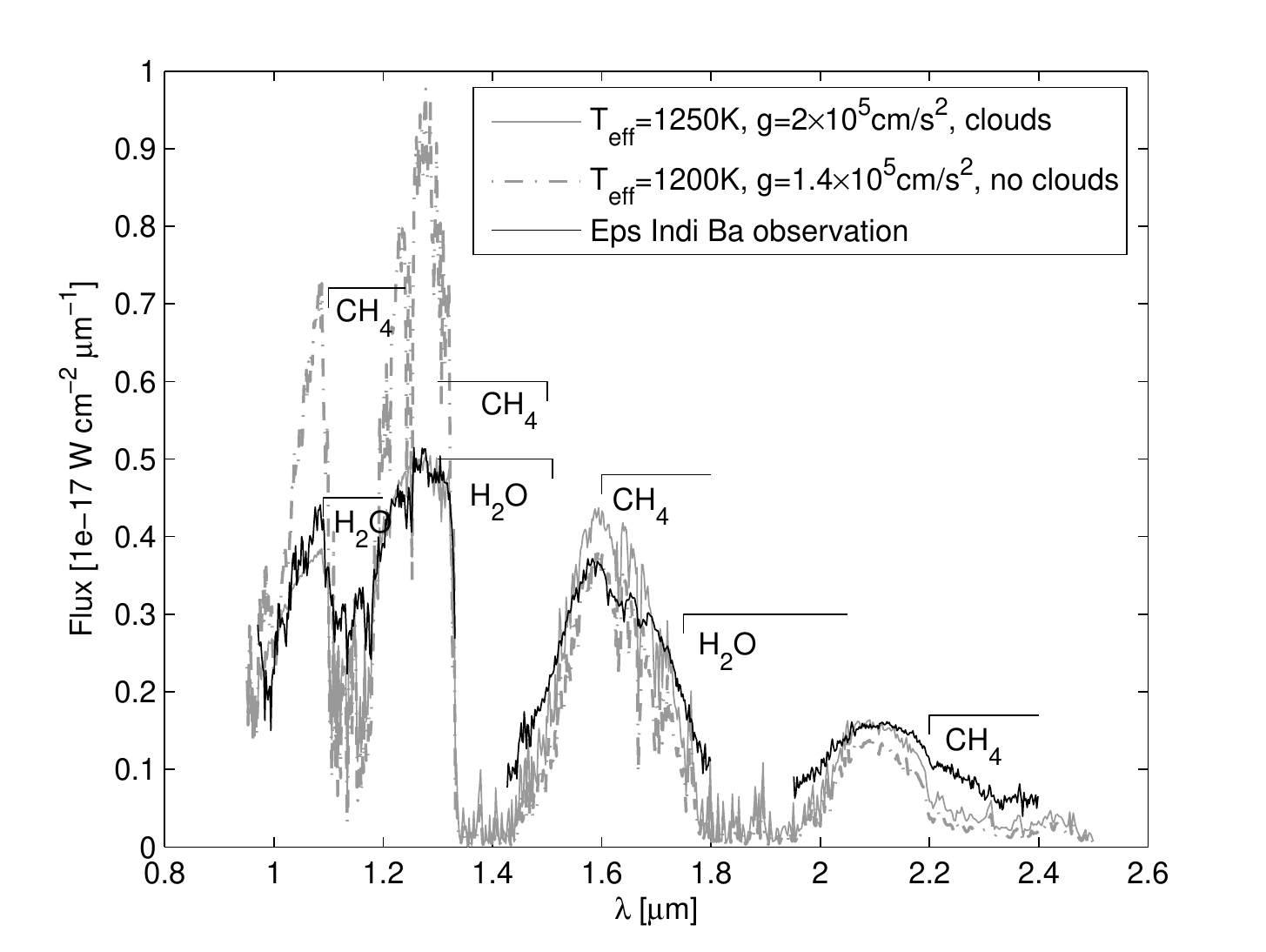}
\caption{NIR spectrum of Eps Indi Ba. The underlying solid gray curve shows the best model fit. The dotted line shows a similar model not considering clouds.}
\label{fig:jhkspectraA}
\end{figure}

The fits to Eps Indi Ba are somewhat less good, but constraining.  Fitting the $Z$ and $J$ bands was compromised by ambiguities in the cloud opacities and K\,I wings. To approximately achieve the observed flux level in the $K$ band required clouds to suppress the fluxes in the $Z$ and $J$ bands. Without cloud opacity, which naturally is most efficacious in these bands, the $Z$ and $J$ fluxes would be too high, while the $K$ band flux would simultaneously be too low as shown in figure\,\ref{fig:jhkspectraA}. Clouds level out these bands and without clouds there are no good fits to Eps Indi Ba. An initial investigation showed that the best fit modal particle sizes were $\sim$$15 \pm 5$ $\mu$m. This is an interesting conclusion, albeit arrived at in lieu of a robust theory of condensates in brown dwarf atmospheres. The three best fitting models are derived for (T$_{\rm eff}$ / log($g$)) of (1250 / 5.3), (1250 / 5.2) and (1300 / 5.3) in order of preference corresponding to masses and ages of (55\,M$_{\rm J}$ / 1.7\,Gyr), (48\,M$_{\rm J}$ / 1.2\,Gyr) and (56\,M$_{\rm J}$ / 1.5\,Gyr). We finally estimate mass and age of Eps Indi Ba of 53$^{+9-7}$ M$_{\rm J}$ and 1.5$^{+0.8-0.5}$ Gyr. The radius of Eps Indi Ba, taken from \citet{burrows97}, is $\sim$0.83 R$_{\rm J}$. Figure\,\ref{fig:jhkspectraA} displays the measured spectrum together with the best fitting model and a plot portraying a model, but without clouds, demonstrating the important role of clouds.

There is a remarkable agreement in the best fitting ages of Eps Indi Ba and Bb of 1.5 Gyr, so the models are consistent with the assumption of coevality. However, the age range over which the models produce reasonable fits is of the order $\pm$0.5 Gyr, so in total the modelling results agree well with the age estimate of Eps Indi A of 1.3$^{-0.5+0.7}$ Gyr by \citet{lachaume99} suggesting overall spectral and evolutionary model consistency. Our mass estimates are somewhat higher than previous estimates (47 and 28 M$_{\rm J}$ for Eps Indi Ba and Bb, respectively, \citet{mccaughrean04}), but still consistent with those considering the error bars.

There is enough leeway in these fits that does not allow us to further constrain the ages and masses beyond this range. Ambiguities in condensate modeling and the age of Eps Indi A and degeneracies along the T$_{\rm eff}/g$ line in the spectral models are the culprits.  Nevertheless, Eps Indi B is the most promising binary brown dwarf system with which to make absolute comparisons and for which astronomers can test the general theory of brown dwarfs. Hence, these data should be a goad to further measurements and a motivation for improved theory.    

\section{Conclusions}

We have presented the 1st J-band as well as new H- and K-band spectra of Eps Indi Ba and Bb. By comparison with theoretical models we were able to derive strong constraints on effective temperature, surface gravity and age of both brown dwarfs. The results confirm the approximate coevality of the stellar and the two brown dwarf components of the Eps Indi system. The suppressed J-band and enhanced K-band flux of Eps Ind Ba indicates that a noticeable cloud layer is still present in a T1.5 dwarf like Eps Indi Ba at an age of $\approx$1.5\,Gyr, whereas for an T6 dwarf like Eps Ind Bb clouds become much less important at the same age as clouds either sink below the photosphere (larger scale heights) or thin out (see, e.g., \citet{cushing08}). Once dynamical mass estimates derived from monitoring of the orbital motion of the system become available, our slightly increased mass predictions (compared to \citet{mccaughrean04}) of 53$^{+9-7}$\,M$_{\rm J}$ and 34$^{\pm5}$\,M$_{\rm J}$ for Eps Indi Ba and Bb, respectively, will provide a crucial test for the validity of our modeling. Recent dynamical mass estimates point indeed towards a significantly higher system mass for the Eps Ind Ba/Bb system \citep{cardoso08}. Hence the Eps Indi system continues to be a benchmark object for testing and calibrating brown dwarf evolutionary and atmospheric models.

\acknowledgments

This research has made use of data products from the Two Mircon All Sky Survey, a joint project of the University of Massachusetts and IPAC/Caltech, funded by NASA and the NSF. We thank Sandy Leggett for providing electronic M and T dwarf spectra through her Web site. AB would like to acknowledge support from the NASA Astrophysics Theory Program under grant NNX08AU16G. WB acknowledges support by the Deutsches Zentrum f\"ur Luft- und Raumfahrt (DLR), F\"orderkennzeichen 50 OR 0401.



{\it Facilities:} \facility{VLT (NACO)}.

\bibliographystyle{aa}



\clearpage

\end{document}